\documentclass[5p,preprint,final, authoryear,twocolumn]{elsarticle}

\usepackage{amssymb}
\usepackage{amsmath}
\usepackage{amsthm}
\usepackage{ latexsym}
\usepackage{graphicx}
\usepackage{amssymb}
\usepackage{times}
\usepackage{enumerate}
\usepackage{amsmath}
\usepackage{amsfonts}
\usepackage{txfonts}
\usepackage{color}
\usepackage{hyperref}

\newcommand{{\bfkappa}}{\mbox{\boldmath${\kappa}$\unboldmath}}
\newcommand{{\bfg}}{\mbox{\boldmath$g$\unboldmath}}
\newcommand{{\bfv}}{\mbox{\boldmath$v$\unboldmath}}
\newcommand{{\bfk}}{\mbox{\boldmath$k$\unboldmath}}
\newcommand{{\bff}}{\mbox{\boldmath$f$\unboldmath}}
\newcommand{{\bfF}}{\mbox{\boldmath$F$\unboldmath}}
\newcommand{{\bfA}}{\mbox{\boldmath$A$\unboldmath}}

\DeclareMathOperator{\sech}{sech}
\DeclareMathOperator{\asech}{asech}
\hypersetup{
  colorlinks = true,
  citecolor = blue,
  urlcolor = black
}

\journal{New Astronomy}

\journal{New Astronomy}

\begin{document}

\begin{frontmatter}

\title{Magnetorotational Instabilities and Pulsar Kick Velocities}

\author[]{Ricardo Heras}
\fntext[ms]{E-mail: ricardo.heras.13@ucl.ac.uk}

\address{Department of Physics and Astronomy, University College London, Gower Street, London WC1E 6BT, United Kingdom}

\begin{abstract}
At the end of their birth process, neutron stars can be subject to a magnetorotational instability in which a conversion of kinetic energy of differential rotation into radiation and kinetic energies is expected to occur at the Alfv\'en timescale of few ms. This birth energy conversion predicts the observed large velocity of neutron stars if during the evolving of this instability the periods are of few ms and the magnetic fields reach values of $10^{16}$G.
\end{abstract}

\begin{keyword}
Neutron Stars, Magnetic Fields, Magnetorotational Instabilities
\end{keyword}

\end{frontmatter}

\noindent A major unsolved problem in astrophysics is to explain why neutron stars exhibit space velocities well above those of their progenitor stars  \citep{anderson:83,lai:01,lai:04,fryer:04,chatterjee:05,hobbs:05,postnov:06}. Most neutron stars have been measured to possess high space velocities in the range of 100-1000 km/s, while their progenitor stars have velocities of the order of 10-20 km/s  \citep{lyne:94,arzoumanian:02,fryer:98,hansen:97,lorimer:97,cordes:98}. It is generally accepted that neutron stars receive a substantial kick at birth, which produce
their observed space velocities. However, the physical origin of this kick is unclear. Some of the proposed kick mechanisms require large initial magnetic fields in the magnetar range ($10^{14}\!-ç\!10^{16}$G) and asymmetric emission of neutrinos \citep{kusenko:96,kusenko:97,lai:98,maruyama:11}. Other mechanisms require a rapid initial rotation to produce substantial kicks \citep{spruit:98,khokhlov:99,sawai:08}. Hydrodynamical models have also been suggested, which are based on recoil due to asymmetric supernovae \citep{burrows:96,scheck:04,burrows:07,nordhaus:12}. Recent models rely on the existence of topological vector currents \citep{charbonneau:10}. The current distribution of the observed neutron star velocities seems to be Maxwellian, which points to a common acceleration mechanism \citep{hansen:97,hobbs:05}. The elusive kick mechanism appears to be connected with the possibility that during some stage of their birth process, neutron stars reach magnetic fields typical of magnetars and periods typical of millisecond pulsars \citep{usov:92,duncan:92,thompson:94}.

In this paper we suggest that the kick velocity of neutron stars may arise from a magnetorotational instability (MRI) produced at the end of their
birth process. More specifically, a newly-born neutron star can be subject to a MRI evolving at the Alfv\'en time of a few ms in which an emission of radiation energy accompanied by a gain of kinetic energy of translation can be produced at the expense of a loss of kinetic energy of differential rotation. If during the evolving of this birth MRI the period is of a few ms and the magnetic field reaches values of $10^{16}$G then we show that the gain of kinetic energy of translation can predict the observed large kick velocity of neutron stars. Our suggestion is supported by studies showing that there is an amplification of the magnetic field and a transference of angular momentum during the evolving of MRI in newly-born neutron stars \citep{akiyama:03,thompson:05,masada:12}. Simulations have shown that birth MRI can generate magnetic fields of the order of $10^{16}\!-\!10^{17}$G in several ms \citep{thompson:05, siegel:13}.

Newly-born neutron stars are assumed to be highly convective and differentially rotating hot fluids \citep{stergioulas:03,yamada:04}, which can be subject to MRI evolving at Alfv\'en timescales \citep{duez:06}. During this birth MRI the star can convert its kinetic energy of differential rotation into magnetic energy \citep{akiyama:03,spruit:08}.
It is then plausible to assume that the star can lose kinetic energy of differential rotation via radiation (a fraction of the total magnetic energy). Conceivable, in this stage the star can also gain kinetic energy of translation and this is the basic assumption of the model proposed here.
We assume that at the end of a birth MRI, neutron stars experience the energy conversion
\begin{align}
{\rm P}_{\rm rad}+\frac{d}{dt}\bigg(\frac{M{\rm v}^2}{2}\bigg)=-\frac{d}{dt}\bigg(\frac{\alpha_{\rm d}I\Omega^2}{2}\bigg),
\end{align}
where P$_{\rm rad}$ is the instantaneous radiated power; $M{\rm v}^2/2$ is the kinetic energy; $\alpha_{\rm d}I\Omega^2/2$ is the kinetic energy of differential rotation; ${\rm v}$ is the space velocity; $\alpha_{\rm d}$ is
a dimensionless constant parameter accounting for the differential rotation \citep{spruit:08}; $I=2MR^2/5$ is the moment of inertia with $M$ and $R$ being the mass and radius; and $\Omega=2\pi/P$ is the angular velocity with $P$ being the period. In Eq.~(1) we have assumed that there are not significant changes of the parameter $\alpha_{\rm d}$ during the short period occurring the birth MRI. But in general $\alpha_{\rm d}$ varies with time.

According to Eq.~(1) an emission of radiation energy and an increase of kinetic energy of translation occur at the expense of a loss of
kinetic energy of differential rotation. We note that an equation similar to Eq.~(1) [without $\alpha_{\rm d}$ and with P$_{\rm rad}$] associated with the asymmetric radiation from an off-centered magnetic dipole is the basis of the ``rocket model'' proposed by \cite{harrison:75}. Another equation also similar to Eq.~(1) [without $\alpha_{\rm d}$ and with P$_{\rm rad}$ estimated by an exponential field decay law] has been considered to study both the birth accelerations of neutron stars \citep{heras:13} and the birth-ultra-fast-magnetic-field decay of neutron stars \citep{heras:12}. It is pertinent to say that the idea that newly-born neutron stars would lose their rotational energy catastrophically on a timescale of seconds or less was suggested by \cite{usov:92}.

The instantaneous radiated power ${\rm P}_{\rm rad}$ in Eq.~(1) must express the idea that a MRI is responsible for a rapid exponential growth of the magnetic field \citep{akiyama:03,yamada:04,siegel:13}. In this sense, \cite{spruit:08} has pointed out that some form of MRI occurring during a differential rotation in the final stages of the core collapse phase may produce an exponential growth of the magnetic field and that once formed, the magnetic field is in risk of decaying again by magnetic instabilities. Without considering changes of kinetic energy, it is well-known that abrupt changes of kinetic energy of rotation
produce abrupt changes of energy of radiation. The characteristic time of rotation changes is similar to the characteristic time of radiation changes. Therefore if the birth MRI occurs at Alfv\'en times of ms then the radiative energy must be emitted in these times and therefore the increase and decrease of magnetic fields producing such a radiative energy must occur at these times. Accordingly, we assume here the existence of a birth MRI responsible for a rapid exponential growth of the magnetic field, followed by an equally rapid exponential decay of this field. More explicitly: at the start of the assumed MRI, the magnetic field has the value $B_{\rm A}$ and then it exponentially grows to reach its maximum value $B_{\rm M}$, followed by a rapid exponential decrease reaching the final value $B_A$. The exponential growth and decay rates of the magnetic field are assumed to occur with the same characteristic time. In a more general treatment, we can assume that the characteristic times of the field increasing is different from that of the field decaying. However, both times must be of the order of Alfv\'en times of ms. A similar field behaviour but for an electric field has been discussed in electromagnetism for the decay of the electric dipole moment \citep{schantz:95}. Expectably, after this abrupt birth field decay, there will be a subsequent field decay caused by Ohmic diffusion and/or other resistive processes, which occur on time scales much larger than those of the birth MRI.

An exponential growth occurs when the growth rate of the function is proportional to the current value of this function: $\dot{f}(t)\!\propto\! f(t)$, where the overdot means time differentiation. Analogously, an exponential decay occurs when $\dot{f}(t)\!\propto\! -f(t)$. For the case of a magnetic moment $\mu(t)$ the positive and negative growth rates can be described by the relation $\dot{\mu}(t)\!\propto \!-\mu(t)\tanh(t/\tau_a)$,
where the hyperbolic tangent function has been introduced to describe the positive and negative growth rates. The time $\tau_a$ is the associated characteristic time. Accordingly, the behaviour of the magnetic moment of a newly-born neutron star during a birth MRI is assumed to be described by the equation $\dot{\mu}(t)\!=\!-(1/\tau_a)\mu(t)\tanh(t/\tau_a)$, whose solution reads
\begin{align}
 \mu(t)\!=\!\mu_{\rm M}\sech \,(t/\tau_a),
 \end{align}
where $\mu_{\rm M}$ is the maximum value of the magnetic dipole moment satisfying $\mu(0)\!=\!\mu_{\rm M}.$
For a neutron star of radius $R$, the magnetic moment $\mu$ is related to the magnetic field $B$ by means of $ \mu(t)\!=\!B(t)R^3$ \citep{jackson:98}, which can be used together with Eq.~(2) to obtain the expected magnetic field law
\begin{align}
 B(t)\!=\!B_{\rm M}\sech \,(t/\tau_a),
\end{align}
where $B_{\rm M}=B(0)$ is the maximum value of the magnetic field. The behaviour of the magnetic field in the assumed birth MRI
is qualitatively shown in Fig.~1.
\begin{figure}[h]
  \centering
    \includegraphics[scale=.22]{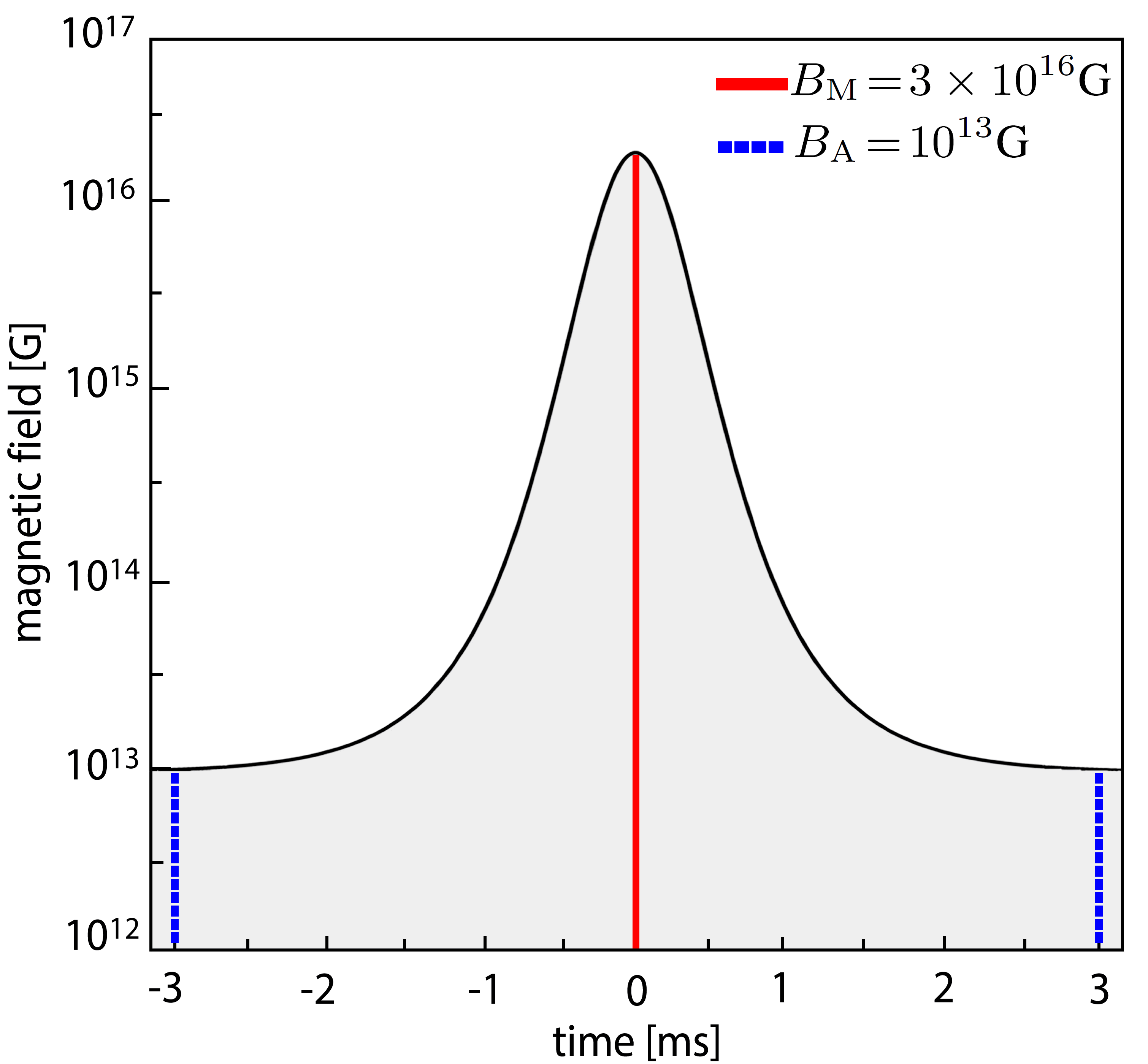}
\caption{Qualitative behaviour of the magnetic field of a newly-born neutron star during the assumed MRI. The
times $-\tau_{\rm A}/2\!=\! -3$ ms and $\tau_{\rm A}/2\!=\!3$ ms correspond to the beginning and end of this birth MRI. The value of the magnetic field at the beginning and end of the MRI is $B_{\rm A}\!=\!10^{13}$G. The maximum value of the magnetic field $B_{\rm M}\!=\!3\times10^{16}$G occurs when $B(0)\!=\!B_{\rm M}.$ }
\end{figure}
\noindent
Equation (2) implies the non-linear equation $\ddot{\mu}^2-4(\dot{\mu}^4/\mu^2-\dot{\mu}^2/\tau_a^2)-\mu^2/\tau_a^4=0$, which combines with
$\mu(t)\!=\!B_{\rm M}R^3\sech \,(t/\tau_a)$ to yield
 \begin{align}
\ddot{\mu}^2(t)=\!\frac{B_{\rm M}^2R^6}{\tau_a^4}\sech\bigg(\!\frac{t}{\tau_a}\!\bigg)^2\bigg[2\tanh\bigg(\!\frac{t}{\tau_a}\!\bigg)^2\!-\!1\bigg]^2.
 \end{align}
Using the Larmor formula ${\rm P}_{\rm rad}(t)\!=\!2\ddot{\mu}(t)^2/(3c^3)$ and Eq.~(4), we obtain the instantaneous radiated power by a newly-born neutron star during  the birth MRI,
\begin{equation}
{\rm P}_{\rm rad}\!=\!\frac{2B_{\rm M}^2R^6}{3c^3\tau_a^4}\sech\bigg(\!\frac{t}{\tau_a}\!\bigg)^2\bigg[2\tanh\bigg(\!\frac{t}{\tau_a}
\!\bigg)^2\!-\!1\bigg]^2.
\end{equation}
The change of kinetic energy of differential rotation can be written as
$d(\alpha_{\rm d} I\Omega^2/2)/dt =(4\alpha_{\rm d}\pi^2MR^2/5)d (1/P^2)/dt.$ Using this relation together with Eqs.~(1) and (5) we obtain the instantaneous energy conversion occurring in the birth MRI:
\begin{eqnarray}
\frac{2B_{\rm M}^2R^6}{3c^3\tau_a^4}\sech\bigg(\!\frac{t}{\tau_a}\!\bigg)^2\Bigg[2\tanh\bigg(\!\frac{t}{\tau_a}\!
\bigg)^2\!\!\!-\!\!1\Bigg]^2\!\!+\!\frac{d}{dt}\Bigg[\!\frac{M{\rm v}^2(t)}{2}\!\Bigg]\nonumber\\
=\!-\frac{4\alpha_{\rm d} \pi^2MR^2}{5}
\frac{d}{dt}\bigg[\!\frac{1}{P^2(t)}\!\bigg].\qquad
\end{eqnarray}
Let us consider the time $\tau_{\rm A}$ during which the MRI evolves. We will call $\tau_{\rm A}$ the MRI time. We adopt the conditions $B(\pm\tau_{\rm A}/2)\!=\!B_{\rm A}$, where $B_{\rm A}$ is the value of the magnetic field at the beginning and end of the birth MRI. These conditions and the law in Eq.~(2) imply the result $B_{\rm A}\!=\!B_{\rm M}\sech\,[\tau_{\rm A}/(2\tau_a)]$ or equivalently
\begin{align}
\tau_{\rm A}\!=\!2\tau_a\asech (B_{\rm A}/B_{\rm M}),
\end{align}
where asech denotes the inverse hyperbolic secant function.

Integration of Eq.~(6) over the time interval in which the assumed birth MRI evolves: $-\tau_{\rm A}/2\leq t\leq\tau_{\rm A}/2$, or equivalently, $-\tau_a\asech\,(B_{\rm A}/B_{\rm M})\leq t\leq\tau_a\asech\,(B_{\rm A}/B_{\rm M})$, yields
\begin{align}
&\underbrace{\frac{4}{45}\frac{R^6(B_{\rm M}^2\!-\!B_{\rm A}^2)^{1/2}(7B_{\rm M}^4\!+\!12B_{\rm A}^4\!-\!4B_{\rm M}^2B_{\rm A}^2)}{B_{\rm M}^3c^3\tau_a^3}}_{\Delta E_{\rm rad}} +\underbrace{\frac{M {\rm v}^2}{2}}_{\Delta E_{\rm kin}}\nonumber\\ &\qquad\qquad\qquad\qquad
=\!\underbrace{\frac{4\alpha_{\rm d}\pi^2MR^2}{5}\bigg(\frac{1}{P_a^2}\!-\!\frac{1}{P_b^2}\bigg)}_{\alpha_{\rm d}\Delta E_{\rm rot}},
\end{align}
where $P_a\!=\!P(-\tau_{\rm A}/2)$ and $P_b\!=\!P(+\tau_{\rm A}/2)$ are the values of the period at the beginning and the end of the birth MRI. The quantity ${\rm v}\!=\!{\rm v}(\tau_{\rm A}/2)$ denotes the space velocity at the end of this instability, which is assumed to be the current value of the observed kick velocity. The initial condition ${\rm v}(-\tau_{\rm A}/2)\!=\!0$ has been adopted, which is consistent with the idea that newly born stars acquire their space velocity during the birth MRI.

If  $B_{\rm M}\!\gg\!B_{\rm A}$ then the radiation term in Eq.~(8) reduces to $\Delta E_{\rm rad}\approx (28/45)R^6B_{\rm M}^2/(c^3\tau_a^3)$
and thus Eq.~(8) becomes
\begin{align}
\frac{28}{45}\frac{R^6B_{\rm M}^2}{c^3\tau_a^3} +\frac{M {\rm v}^2}{2}=\!\frac{4\alpha_{\rm d}\pi^2MR^2}{5}\bigg(\frac{1}{P_a^2}\!-\!\frac{1}{P_b^2}\bigg).
\end{align}
We expect that the time $\tau_{\rm A}$ is of a few ms for magnetic fields of the order of $10^{16}$G. This expectation restricts the values of the time $\tau_a$ and those of the relation $B_{\rm A}/B_{\rm M}$. In particular, if $\tau_a$ is taken to be $\tau_a\!\approx \!10 R/c$
then Eq.~(7) implies a MRI time $\tau_{\rm A}$  in the interval $ 3\;{\rm ms}\lesssim\tau_A\!\lesssim  6\,$ms when $B_{\rm M}\!=\!3\times\! 10^{16}$G and $B_{\rm A}$ lies in the interval $10^{13}{\rm G}\!\leq\! B_{\rm A}\!\leq\! 10^{15}{\rm G}$. The choice $\tau_a\approx 10 R/c$ is then consistent with the model proposed here. Insertion of $\tau_a\!\approx\! 10 R/c$ into Eq.~(9) yields
\begin{align}
\frac{7R^3B_{\rm M}^2}{11250} +\frac{M {\rm v}^2}{2}=\!\frac{4\alpha_{\rm d}\pi^2MR^2}{5}\bigg(\frac{1}{P_a^2}\!-\!\frac{1}{P_b^2}\bigg).
\end{align}
Let us emphasize that Eq.~(10) is valid when the magnetic field $B_{\rm M}$ is given by Eq.~(3) and the conditions $B_{\rm M}\!\gg\!B_{\rm A}$ and $\tau_a\approx 10R/c$ are fulfilled. Equation (10) can be used to constrain the values of the periods $P_a$ and $P_b$. In fact, Eq.~(10) with $\alpha_{\rm d}\!=\!0.1, M\!=\!1.4 M_\odot$ and $R\!=\!10$ km imply
\begin{align}
P_b=\frac{1}{\sqrt{P_a^{-2}-(\kappa_1B_{\rm M}^2+\kappa_2{\rm v}^2)}},
\end{align}where $\kappa_1\!\approx\! .281\!\times\! 10^{-30}$ cm/gr and $\kappa_2\!\approx\! .633\!\times\! 10^{-12}$\,cm$^{-2}$. Since $P_b$ is real it follows that
 \begin{align}
P_a\!<\frac{1}{\sqrt{\kappa_1B_{\rm M}^2+\kappa_2{\rm v}^2}}.
\end{align}
This condition restricts the values of $B_{\rm M}$ and $ {\rm v}$.
By assuming $B_{\rm M}\!=\!3\times\! 10^{16}$G, we will apply Eqs.~(11) and (12) to three different velocities: (a) If ${\rm v}\!=\!100$ km/s then Eq.~(12) implies $P_a\!<\!.0561$ s. In particular, if $P_a\!=\!.015$ s then Eq.~(11) gives $P_b\!\approx.0155$ s; (b)
If ${\rm v}\!=\!500$ km/s then $P_a\!<\!.0233$ s. For example, if $P_a\!=\!.01$ s then $P_b\!\approx\! .0108$ s; and (c) If ${\rm v}\!=\!1000$ km/s then $P_a\!<\!.0123$ s. For instance, if $P_a\!=\!.005$ s then $P_b\!\approx \!.0055$ s. If the initial period satisfies $P_a\!\leq\!.015$ s then the changes $\Delta P\!=\!P_b\!-\!P_a$ (for the previous examples) are of the order of $10^{-4}$ s for the interval of space velocities $100$ km/s $\leq{\rm v}\leq 1000$ km/s.

We can use Eq.~(10) to obtain a formula for the kick velocity
 \begin{align}
{\rm v}_{\rm kick}\!=\sqrt{\frac{8\alpha_{\rm d}\pi^2R^2}{5}\bigg(\frac{1}{P_a^2}\!-\!\frac{1}{P_b^2}\bigg)- \frac{7R^3B_{\rm M}^2   }{5625 M}   },
\end{align}
where we have written ${\rm v}\!\equiv\!{\rm v}_{\rm kick}$ to emphasize that the kick velocity of neutron stars is originated by the birth MRI quantities: $\alpha_{\rm d}, B_{\rm M}, P_a$ and $P_b$. This result explains
the absence of the suspected correlation between the current magnetic field and space velocity, the so-called ``${\rm v}\!-\!B$ correlation''  \citep{lorimer:95,lorimer:97,cordes:98}. Let us rewrite Eq.~(13) as
 \begin{align}
{\rm v}_{\rm kick}\!=\sqrt{{\rm v}^2_{[\rm drot]}\!-\! {\rm v}^2_{[\rm rad]} }\;,
\end{align}
where ${\rm v}_{[\rm drot]}$ is the component of the kick velocity originated by the change of kinetic energy
of differential rotation
 \begin{align}
{\rm v}_{[\rm drot]}=\pi R\sqrt{\frac{8\alpha_{\rm d}}{5}\bigg(\frac{1}{P_a^2}\!-\!\frac{1}{P_b^2}\bigg)}\;,
\end{align}
and ${\rm v}_{[\rm rad]}$ is the component of the kick velocity originated by the change of radiation energy
 \begin{align}
{\rm v}_{[\rm rad]}=\frac{B_{\rm M}}{75}\sqrt{\frac{7R^3}{M}}\;.
\end{align}
We can infer the kick force ${\rm F}_{\rm kick}(t)$ producing ${\rm v}_{\rm kick}(t)$. Time-integration of Eq.~(6) from  $-\tau_A/2$ to the time $t$ leads to
\begin{align}
{\rm v}_{\rm kick}(t)\!=\sqrt{{\rm v}^2_{[\rm drot]}(t)\!-\! {\rm v}^2_{[\rm rad]}(t)},
\end{align}
where
\begin{align}
{\rm v}_{[\rm drot]}(t)=\sqrt{\kappa\Bigg[\frac{1}{P_a^{2}}\!-\!\frac{1}{P(t)^{2}}\Bigg]}\;,
\end{align}
in which $\kappa= (8\alpha_{\rm d}\pi^2R^2)/5, P_a\!=\!P(-\tau_{\rm A}/2),$ and
\begin{align}
{\rm v}_{[\rm rad]}(t)=\sqrt{\frac 2M\int_{-\tau_A/2}^t\!\! {\rm P}_{\rm rad}(\sigma)d\sigma }\;.
\end{align}
If Eq.~(18) is evaluated in the time $t=\tau_A/2$ and we write $P_b\!=\!P(\tau_{\rm A}/2)$ then we obtain Eq.~(15). Analogously, if Eq.~(19) with ${\rm P}_{\rm rad}$ given by Eq.~(5) is evaluated in $t=\tau_A/2$ and the conditions $B_{\rm M}\!\gg\!B_{\rm A}$ and $\tau_a\approx 10R/c$ are fulfilled then we obtain
Eq.~(16). Using Eqs.~(18) and (19) we can derive the corresponding accelerations. They can be expressed as
\begin{align}
{\rm a}_{[\rm drot]}(t)=\frac{\kappa\dot P(t)}{P^3(t){\rm v}_{[\rm drot]}(t)},\;\;{\rm a}_{[\rm rad]}(t)=\frac{{\rm P}_{\rm rad}(t)}{{\rm v}_{[\rm rad]}(t)}.
\end{align}
Time differentiation of Eq.~(17) and the use of Eqs.~(20) gives the kick (total) acceleration: ${\rm a}_{[\rm kick]}={\rm v}_{[\rm drot]}{\rm a}_{[\rm drot]}/{\rm v}_{[\rm kick]}-{\rm v}_{[\rm rad]}{\rm a}_{[\rm rad]}/{\rm v}_{[\rm kick]}$, which can alternatively be expressed as
\begin{align}
{\rm a}_{[\rm kick]}(t)=\frac{\kappa {\dot P}(t)}{P(t)^3{\rm v}_{[\rm kick]}(t)}-\frac{{\rm P}_{\rm rad}(t)}{M{\rm v}_{[\rm kick]}(t)}.
\end{align}
We can evaluate ${\rm a}_{[\rm kick]}$ at $t\!=\!0$ to get an idea of the order of its magnitude. According to Eq.~(5), the power radiated at $t\!=\!0$ is ${\rm P}_{\rm rad}(0)\!\approx \!.1874\times 10^{52}$ ergs for $B_{\rm M}\!=\!3\times 10^{16}$G, $R\!=\!10$ km and $\tau_a\!=\!3.33\times 10^{-4}$s. From Eq.~(17) it follows that ${\rm v}_{[\rm kick]}(0)\!\approx\! 128$ km/s if we assume $P_a\!=\!.019$ s, $P(0)=.01986$ s and $M\!=\!1.4 M_\odot$. We write $\kappa\!\approx\!.158 \times 10^{13}$ cm$^2$ (for $\alpha_{\rm d}\!=\!0.1$ and $R\!=\!10$ km) and ${\dot P}(0)\!\approx\! 6.62$ (calculated by ${\dot P}\!\sim \!2P/\tau_A$). Using all of these values in Eq.~(21) we get: ${\rm a}_{[\rm kick]}(0)\!\approx\! .53 \times 10^8$ g \citep{heras:13}.

The forces associated with accelerations in Eqs.~(20) read
\begin{align}
{\rm F}_{[\rm drot]}(t)=\frac{M\kappa\dot P(t)}{P^3(t){\rm v}_{[\rm drot]}(t)},\;{\rm F}_{[\rm rad]}(t)=\frac{M{\rm P}_{\rm rad}(t)}{{\rm v}_{[\rm rad]}(t)}.
\end{align}
By making use of ${\rm F}_{[\rm kick]}\!\!=\!\!({\rm v}_{[\rm drot]}{\rm F}_{[\rm drot]}\!\!-\!\!{\rm v}_{[\rm rad]}{\rm F}_{[\rm rad]})/{\rm v}_{[\rm kick]}$ and Eqs.~(22) we can obtain the kick (total) force
\begin{align}
{\rm F}_{[\rm kick]}(t)=\frac{\kappa M{\dot P}(t)}{P(t)^3{\rm v}_{[\rm kick]}(t)}-\frac{{\rm P}_{\rm rad}(t)}{{\rm v}_{[\rm kick]}(t)}.
\end{align}
It should be noted that Eq.~(23) can naturally be interpreted in the context of the radiation reaction theory. If $M{\rm a}={\rm F}_{[\rm kick]}; {\cal F}_{\rm ext}=\kappa M{\dot P}/(P^3{\rm v}_{[\rm kick]})$; and ${\cal F}_{\rm rad}\!=\!-{\rm P}_{\rm rad}/{\rm v}_{[\rm kick]}$ then we obtain the well-known equation for the radiation reaction \citep{jackson:98}: $M{\rm a}={\cal F}_{\rm ext}+ {\cal F}_{\rm rad}.$

The velocity v$_{[\rm rad]}$ in Eq.~(16) can be related to the Alfv\'en velocity ${\rm v}_{[\rm Alf]}\!= \!B/\sqrt{4\pi\rho}$. For a sphere of mass $M$ and radius $R$ the Alfv\'en velocity reads ${\rm v}_{[\rm Alf]}\!=\! B/\sqrt{3M/R^3}$, which combines with Eq.~(16) (making $B\!=\!B_{\rm M}$) to yield
${\rm v}_{[\rm rad]}\!=\!\sqrt{21}{\rm v}_{[\rm Alf]}/75.$ This implies v$_{[\rm Alf]}\!\approx \!16.37{\rm v}_{[\rm rad]}$. If ${\rm v}_{[\rm rad]}\!\approx\! 200$ km/s for $B_{\rm M}\!=\!3\!\times\! 10^{16}$G, $M\!= \!1.4 M_\odot$ and $R\!=\!10$ km then ${\rm v}_{[\rm Alf]}\!\approx \!3.27\times 10^{8}\;{\rm cm/s},$ and this can be combined with ${\rm v}_{[\rm Alf]}\!=\!2R/T_A$ (considering $R\!=\!10$ km) to get the expected Alfv\'en time $T_A\!\approx\!6$ ms, which emphasizes the consistency of the model proposed here.

Let us apply Eqs.~(14)-(16) to two representative neutron stars: The Crab pulsar B0531+21 and the magnetar  J1809-1943. The canonical values $M\!= \!1.4 M_\odot$ and $R\!=\!10$ km will be assumed.
Unless otherwise specified, all data is taken from the ATNF Pulsar Catalogue \citep{manchester:05}. The space velocity ${\rm v}$ is obtained from the transverse velocity ${\rm v}_{\perp}$ by means of ${\rm v}\!\approx\!\sqrt{3/2}\:{\rm v}_{\perp}$
\citep{hobbs:05,lyne:94}.

Consider first the Crab pulsar with its current transverse velocity $ {\rm v}_{\perp}\!=\!141$ km/s, which implies the space velocity ${\rm v}\!\approx\!172$ km/s.  It has been suggested that this pulsar was born with a period of 19 ms \citep{lyne:93}. If during a birth MRI in the Crab pulsar there was a very small increase in the period, for example, from $P_a\!=\!.019$ s to $P_b\!=\!.02072$ s and an abrupt change of magnetic field starting with the value $10^{13}$G, reaching the maximum value $3\!\times\! 10^{16}$G and ending with the value $10^{13}$G, then Eq.~(14) with $\alpha_{\rm d}=0.1$ yields ${\rm v}_{\rm kick}\!\approx\!172$ km/s and Eqs.~(15) and (16) respectively give ${\rm v}_{[\rm drot]}\!\approx\!264$ km/s and ${\rm v}_{[\rm rad]}\!\approx \!200$ km/s. Notice that the difference between the periods $P_a$ and $P_b$ is small: $\Delta P\!\approx .001$ s. The current period of the Crab pulsar is $P\!=\!.033$ s. Crab's period has increased about 13 ms since the occurrence of the assumed birth MRI.

Consider now the magnetar  J1809-1943, whose current period and magnetic field are $P\!=\!5.54$ s and $B=2.1\times 10^{14}$G. This magnetar has ${\rm v}_{\perp}\!=\!229$ km/s, which implies ${\rm v}\!\approx\!278$ km/s. If this magnetar experienced a birth MRI in which there was a very small increase of its period, for example,  from the value $P_a\!=\!.010$ s to $P_b\!=\!.01039$ s and an abrupt change of magnetic field starting with the value $10^{15}$G, reaching the maximum value $3\!\times\! 10^{16}$G and ending with the value $10^{15}$G, then Eq.~(14) with $\alpha_{\rm d}=0.1$ yields ${\rm v}_{\rm kick}\!\approx\!278$ km/s and Eqs.~(15) and (16) respectively give ${\rm v}_{[\rm drot]}\!\approx\!342$ km/s and ${\rm v}_{[\rm rad]}\!\approx\! 200$ km/s.

It is interesting to note that Eq.~(11) allows the final period $P_b$ to have values considerably larger than those previously assumed. This possibility could explain why some neutron stars exhibit current periods in the scale of seconds (e.g., magnetars). According to Eq.~(11)
the period $P_b$ may be of the same order of the current period $P$. Consider, for example, the magnetar SGR 1806-20 whose current period and magnetic field are $P\!=\!7.54$\,s and $B\!=\!2\!\times\! 10^{15}$G.  This magnetar has  v$_{\perp}\!\approx\!350$ km/s \citep{olausen:14} which implies v $\approx\!428$ km/s.  If we assume the birth MRI values $B_{\rm A}\!=\!4\times\! 10^{15}$G and $P_a\!=\!.0265996$\,s then Eq.~(11) implies $P_b\!=\!6.01$\,s. This result suggests that shortly after the evolving of the assumed birth MRI, this magnetar acquired a period of the order of seconds. However, this conclusion faces the problem of how to physically explain the birth MRI change $\Delta P\!=\!P_b-P_a$ of the order of seconds.

In summary, any convincing explanation for the observed large space velocity of neutron stars should be traced to physical processes occurring during their
birth. Accordingly, the space velocity should be related with birth values of physical properties of these stars and not with current values of them.  Here we have suggested that a MRI produced at the end of the birth process of neutron stars can be responsible for their
large space velocities. We have shown how a rapid birth MRI conversion of kinetic energy of differential rotation into radiation energy and kinetic energy of translation occurring at the Alfv\'en time of a few ms yields an observed interval for neutron star velocities, ranging from several hundreds to a few thousands km/s.


\end{document}